# Frequency Response Assessment of U.S. Power Grids in High PV Penetration

Shutang You

*Abstract*— **Nonsynchronous generations such as photovoltaics (PVs) are expected to undermine bulk power systems (BPSs)' frequency response at high penetration levels. Though the underlying mechanism has been relatively well understood, the accurate assessment and effective enhancement of the U.S. interconnections' frequency response under extra-high PV penetration conditions remains an issue. In this paper, the industry-provided full-detail interconnection models were further validated by synchrophasor frequency measurements and realistically-projected PV geographic distribution information were used to develop extra-high PV penetration scenarios and dynamic models for the three main U.S. interconnections, including Eastern Interconnection (EI), Western Electricity Coordinating Council (WECC), and Electric Reliability Council of Texas (ERCOT). Up to 65% instantaneous PV and 15% wind penetration were simulated and the frequency response change trend of each U.S. interconnection due to the increasing PV penetration level were examined. Most importantly, the practical solutions to address the declining frequency response were discussed. This paper will provide valuable guidance for policy makers, utility operators and academic researchers not only in the U.S. but also other countries in the world.**

## I. Introduction

Nonsynchronous renewable generations, such as wind and photovoltaic (PV) power, have witnessed a rapid growth around the globe [1-5]. For example, on January 7$^{th}$, 2015, the output from Ireland's wind turbine generators (WTG) peaked at 63% of Ireland's total electricity demand. As another example, major economies in the world, e.g. the United States (U.S.), Europe, and China, have installed tens of Giga Watts' WTGs and PVs so far and are expected to install more due to the increasing societal pressure of climate change. Particularly, the U.S. Department of Energy (DoE) SunShot vision study estimated that solar generation may provide up to 14% of the nation's total electricity by 2030 and 27% by 2050. To achieve the 14% and 27% average level, PV may peak at 70% or higher of the total generation at certain moments assuming an average 0.2 capacity factor. At such an extra-high instantaneous penetration rate, the lack of system inertia and decline of frequency response will become a severe issue for bulk power systems (BPSs), such as the three U.S. interconnections, which had been traditionally considered to be less likely affected due to their large sizes.

Computer simulation is the only feasible approach to look into this issue. However, no previous research had ever succeeded in simulating the extra-high PV penetration rates predicted by the DoE SunShot vision study for any of the three U.S. interconnections. Although some recent studies preliminarily revealed the U.S. interconnections' low inertia and decreasing frequency stability issues due to increasing wind or PV penetration, the highest renewable penetration rate in these studies was no more than 50%. Furthermore, some of their models were not validated by measurements or the unbalanced solar power generation distribution was not properly took into consideration, which undermined these studies' prediction credibility. Therefore, in order to accurately assess, not just speculate on, the negative impacts of future extra-high PV penetrations on U.S. interconnections' frequency response, dynamic simulations based on the measurement-validated full-detail interconnection models and realistic PV geographic distribution information will be conducted in this paper.

The major contributions and innovations of this paper include:

1. The three main interconnections in the U.S., including Eastern Interconnection (EI), Western Electricity Coordinating Council (WECC), and Electric Reliability Council of Texas (ERCOT), were all studied in this paper and high-resolution frequency measurements from a wide area monitoring system FNET/GridEye were used to validate the three interconnections' dynamic models at different locations.
2. Extra-high instantaneous PV penetration rates were simulated in this paper, which have never been achieved by previous studies. Furthermore, realistically-projected PV geographic distribution information were utilized in this paper's scenario construction, which added another layer of validity to this study.
3. Practical mitigation solutions to deal with extra-high PV penetration rates were examined in this paper. The effectiveness of changing governor droop, decreasing governor deadbands, and fast load response in improving frequency response were evaluated via simulations.
4. Last but not least, a technical review committee (TRC) that consists of tens of industry experts from the three interconnections provided numerous suggestions to this study in order to make its results as realistic as possible. This work will be extremely valuable to other BPSs which are expected to

Email: syou3 @utk.edu.

reach extra-high PV penetration rates.

The rest of this paper is structured as follows: Section II introduces the process of extra-high PV penetration scenario development and dynamic model construction; Section III simulates each interconnection's frequency response at different PV penetration levels; Section IV discusses the practical solutions available to improve the U.S. interconnections' frequency response; and Section V concludes the entire paper.

## II. Scenario Development and Model Construction

As mentioned earlier, there exist no simulation scenarios or dynamic models for any of the three major U.S. interconnections that meet and exceed the extra-high PV penetration prediction by the DoE SunShot vision study. In this section, the scenario development and model construction efforts will be introduced, including base model validation, scenario determination, PV geographic location projection, and dynamic model construction.

### *A. Study Systems and Base Models*

The EI is one of the largest BPSs in the world and it reaches from Central Canada eastward to the Atlantic coast (excluding Quebec), south to Florida, and back west to the foot of the Rockies (excluding most of Texas). It had a total internal demand of 609 GW in 2015. The multi-regional modeling working group (MMWG) of Eastern Interconnection reliability assessment group (ERAG) built and maintains a single model for the whole EI system, which includes more than 68,000 buses and over 8,000 generators. This full-detail MMWG model will be used as the base model of the EI in this study.

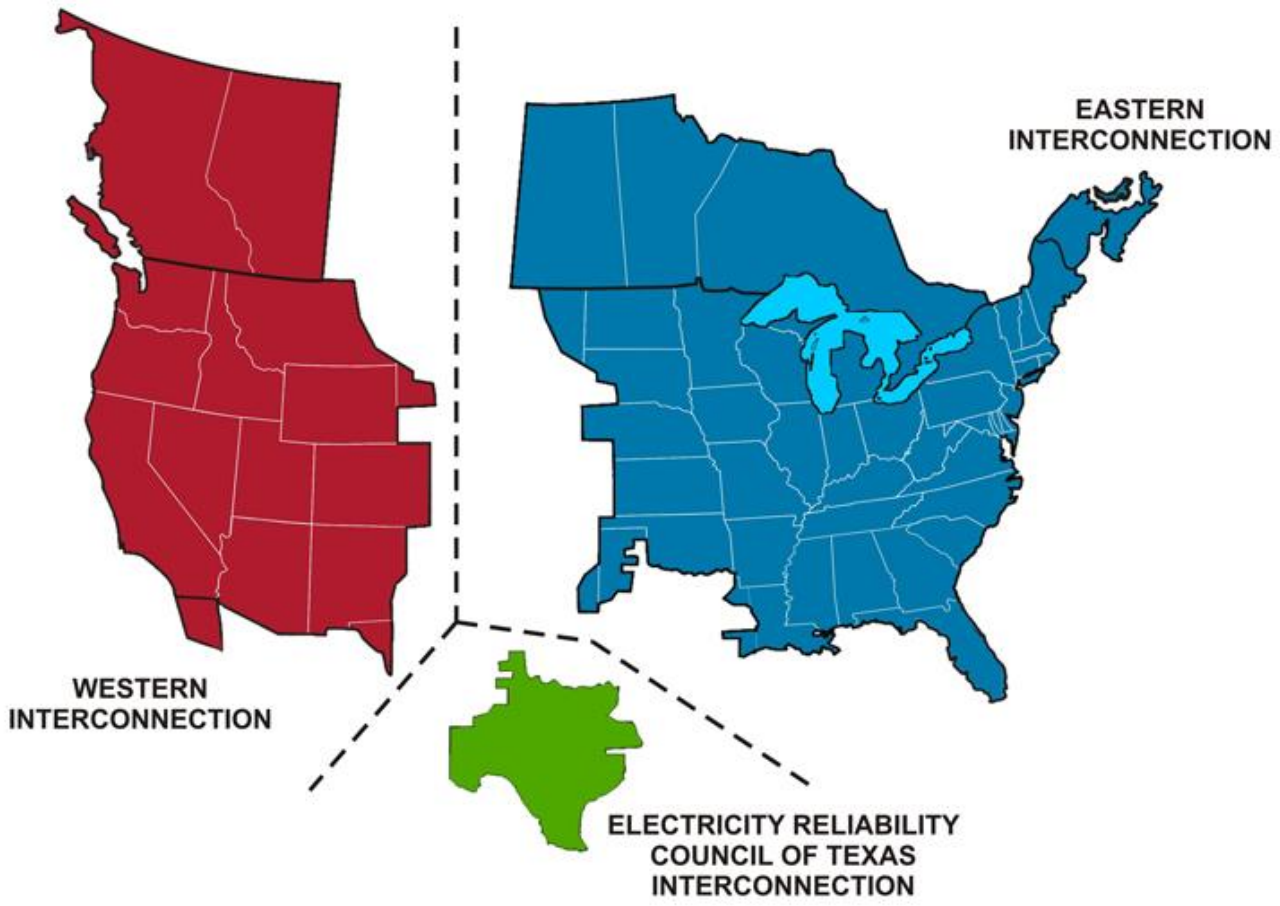

Fig. 1 Map of the three U.S. interconnections **Error! Reference source not found.**

Serving a population of over 80 million, the WECC stretches from Western Canada south to Baja California in Mexico, reaching eastward over the Rockies to the Great Plains. It had a summer peak demand of 150. 7 GW and a winter peak demand of 126.2 GW in 2015. The WECC planning models were assembled annually by the WECC system review group and approved by all the WECC utility members. A WECC planning model typically consists of about 18,000 buses, including over 4,000 generators. Representing 85 percent of the Texas state's electric loads, ERCOT dispatches power on an electric grid that connects 40,500 miles of transmission lines and supply 24 million customers. Power demands in the ERCOT region are highest in summer, primarily due to air conditioning use in homes and businesses. The ERCOT region's all-time record peak hour occurred on August 10, 2015, when consumer demand hit 69.877 GW. The ERCOT model consists of over 6,000 buses, around 700 of which are generators. The geographic locations of the three interconnection are shown in Fig. 1.

### *B. Base Model Validation*

It had been noticed long ago that the widely-used EI MMWG models cannot reflect true EI frequency response accurately. If extra-high PV penetration dynamic models are built on these inaccurate base models, the potential risks posed by PV penetrations may be concealed. Thus the very first step of this study was to tune and validate all the base models based on accurate frequency response measurements. The wide area monitoring system FNET/GridEye has been monitoring the U.S. interconnections for more than a decade and large amounts of frequency measurements have been collected. In this subsection, the FNET/GridEye frequency measurements are leveraged to tune and validate the base models. Please note that the FNET/GridEye frequency measurement accuracy is within ±0.0005 Hz, which is more than accurate enough for frequency response model validation.

According to a previous sensitivity study, total system inertia, governor ratio/spinning reserve, and governor deadband were the major factors that contributed to the base models' inaccurate responses. Therefore, three measures were taken to improve the base models' credibility. Firstly, since the existing base models fail to model governor deadband, realistic governor deadband values were added to the base models. Secondly, the total system inertia values were adjusted to the exact value collected and provided by North American Electric Reliability Corporation (NERC). Thirdly, governor ratio/spinning reserve of the base models were tuned to match the FNET/GridEye frequency measurements of certain contingency. One example of such contingencies in the EI is given in Fig. 2 (a) and the mismatches of frequency response metrics between simulation and measurement, including nadir, rate of change of frequency (ROCOF), and settling frequency, are calculated in Table I. Both Fig.2 (a) and Table I demonstrate that after the three model tuning measures, the EI MMWG model can reflect the actual EI frequency response accurately.

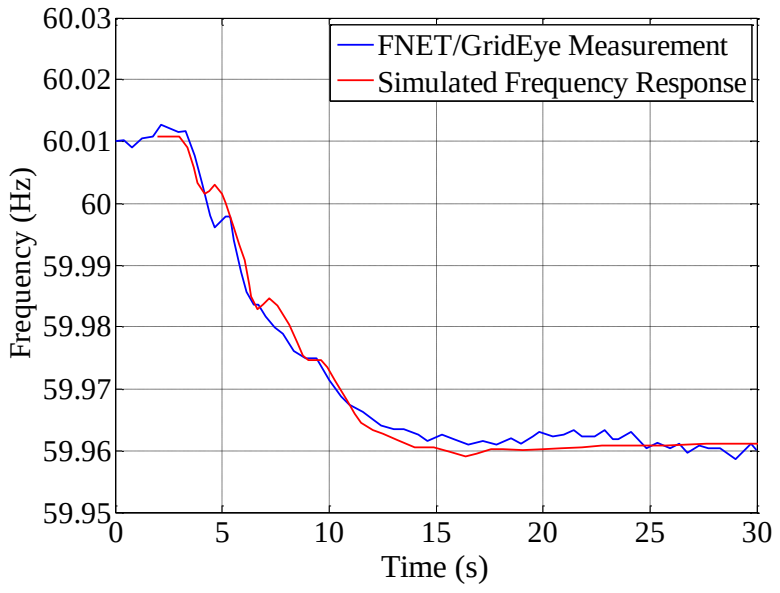

(a) EI

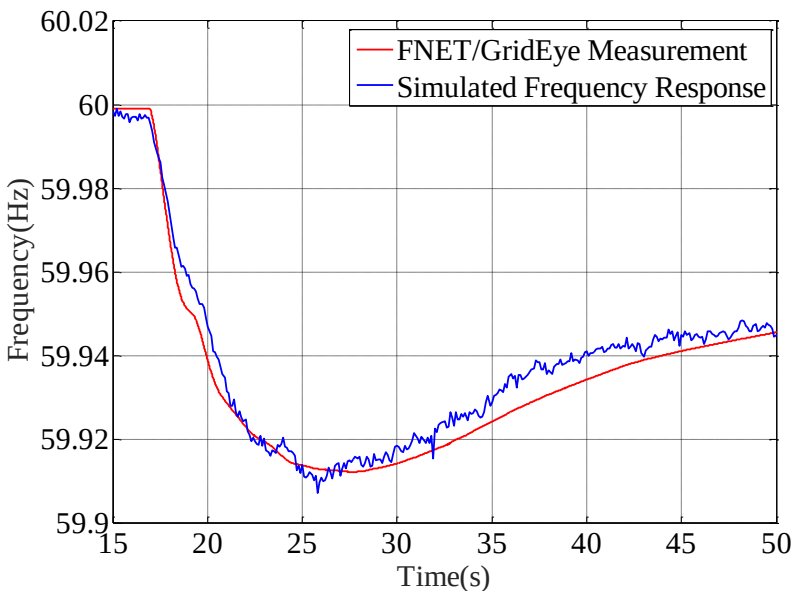

(b) WECC

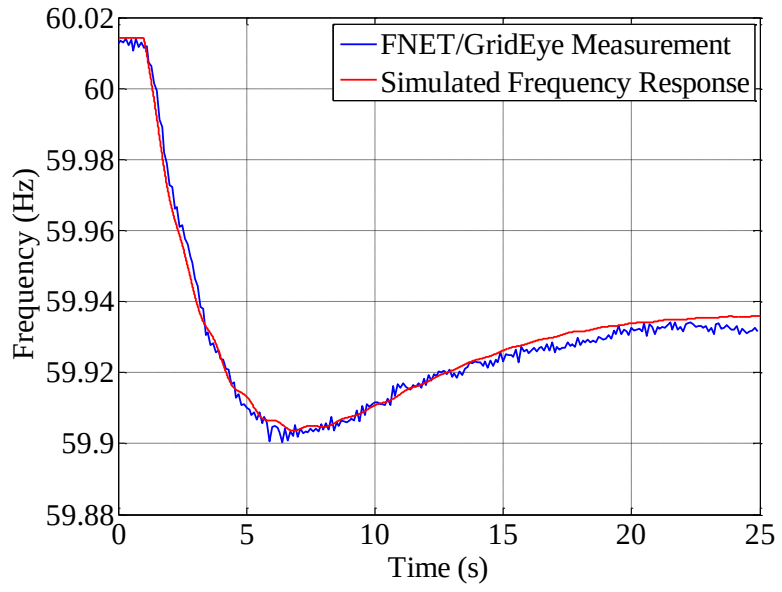

(c) ERCOT

Fig.2 Examples of model validation result

TABLE I
Mismatches of Frequency Response Metrics after Model Validation

| | | Measure -ment | Simulated Value | Error |
|---|---|---|---|---|
| EI | Frequency Nadir (Hz) | 59.9590 | 59.9611 | 0.0021 |
| | ROCOF (mHz/s) | 12.9 | 11.0 | 1.9 |
| | Settling Frequency (Hz) | 59.9618 | 59.9607 | 0.0011 |
| WECC | Frequency Nadir (Hz) | 59.9123 | 59.9133 | 0.0010 |
| | ROCOF (mHz/s) | 9.5 | 8.2 | 1.3 |
| | Settling Frequency (Hz) | 59.9453 | 59.9440 | 0.0013 |
| ERCOT | Frequency Nadir (Hz) | 59.9021 | 59.9036 | 0.0015 |
| | ROCOF (mHz/s) | 43.7 | 45.2 | 1.5 |
| | Settling Frequency (Hz) | 59.9352 | 59.9326 | 0.0026 |

It is worth noting that, the tuned base model's credibility had been checked at multiple locations using multiple contingencies, all of which proved the tuned model's accurate representation of the true EI frequency response. Similar to the EI effort, the ERCOT base model was tuned and validated following the same process and the validation results also verified the tuned ERCOT base model's credibility, as shown in Fig.2 (c) and Table I. Differently, WECC had spent tremendous efforts to tune and validate its models in the past twenty years. As shown in Fig. 2(b), the frequency response produced by the WECC base models can match the FNET/GridEye measurements well without additional tuning measures.

## C. *Scenario Determination*

As discussed earlier, extra-high PV penetration simulation scenarios have to be constructed. Besides, it is also beneficial to build several other simulation scenarios with gradually increasing PV penetration levels to demonstrate the incremental PV impact. To truly reflect the industry needs and maximize the study benefits, a survey was conducted within this study's technical review committee (TRC) about the PV penetration rate in each scenario. This TRC consists of industry experts from major electric utilities such as Dominion Virginia Power (DVP) and Pacific Gas and Electric (PG&E), independent system operators (ISO) such as California ISO and Midcontinent Independent System Operator (MISO), reliability coordinator (RO) such as PEAK Reliability and ERCOT, and other consulting companies and research institutes. Basic assumptions of the scenario selection included: PV generation cost will be significantly lower than that of WTG in the future; nuclear power plants will gradually retire from the generation fleet, and hydro plants will stay for at least two more decades etc. Most importantly, the highest instantaneous PV penetration should be high enough to meet and exceed the PV penetration prediction by the DoE SunShot vision study. According to the survey results, four simulation scenarios were eventually selected by TRC members as presented in Table II. From Table II, up to 65% PV and 80% nonsynchronous renewable penetration rates will be studied, which meets the DoE SunShot vision study prediction.

It should be pointed out that though the same four scenarios were defined for the three interconnections, it does not necessarily indicate they will reach the same penetration level simultaneously. In fact, ERCOT is expected to reach high PV penetration levels much earlier than EI and WECC because of its smaller size and higher solar irradiance.

TABLE II
PV and Renewable Penetration Rates of All Scenarios

| Scenario | Instantaneous PV Penetration | Instantaneous WTG Penetration | Total Renewable (PV and WTG) Penetration |
|---|---|---|---|
| # 1 | 5% | 15% | 20% |
| # 2 | 25% | 15% | 40% |
| # 3 | 45% | 15% | 60% |
| # 4 | 65% | 15% | 80% |

## D. *PV Geographic Location Projection*

Since existing studies did not provide the future PV generation geographic distributions at high penetration levels, they have to be projected reasonably for each interconnection. In this subsection, different approaches were employed for different interconnections' PV distribution projection because of the different data sources available.

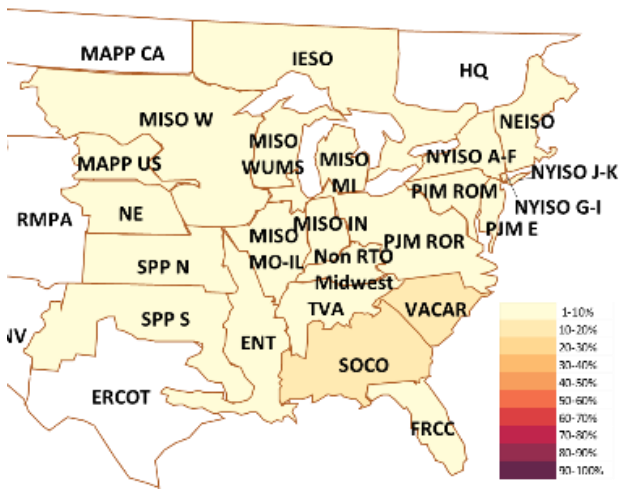

(a) 5% PV

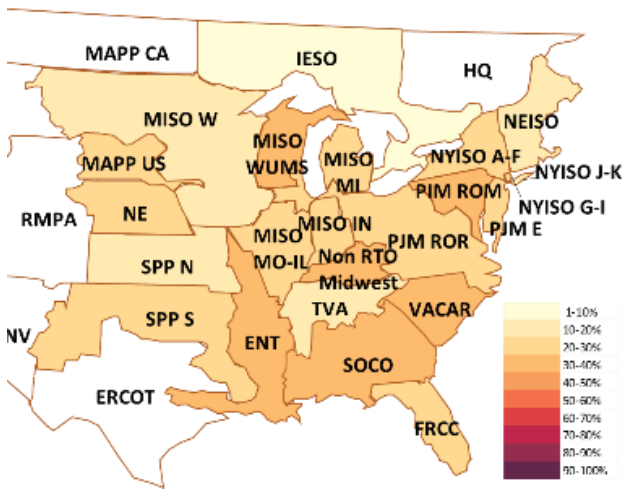

(b) 25% PV

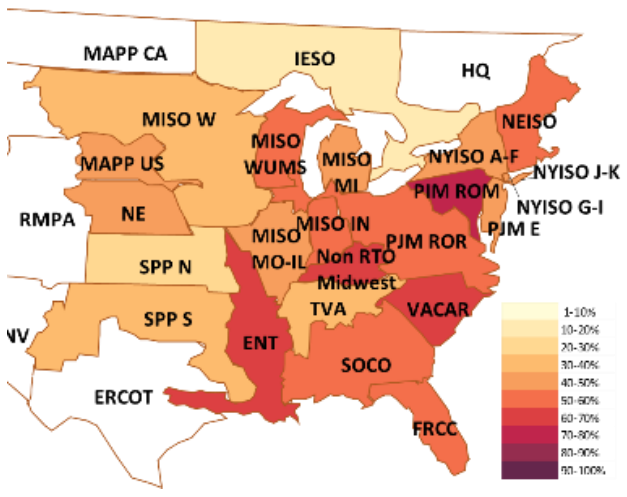

(c) 45% PV

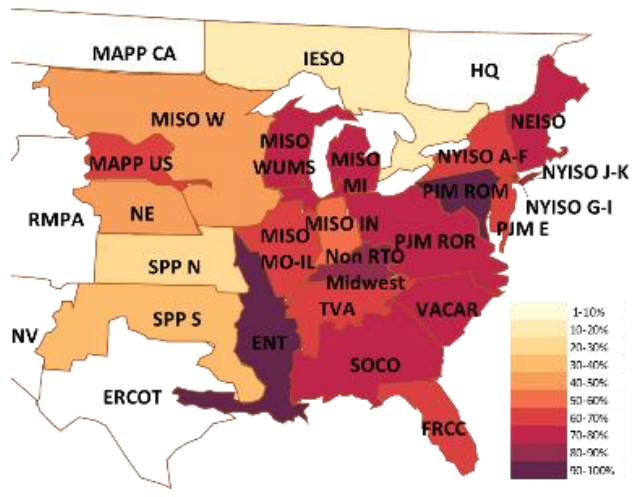

(d) 65% PV

Fig.3 PV penetration rate for each EI region in different simulation scenarios

The future EI PV distribution was calculated by PV expansion optimization in this study, which minimized the system total cost consisting of PV expansion costs, system operation and maintenance costs, and emission costs. The optimization horizon featured an interconnection-level PV growth primarily driven by high carbon-emission prices and PV generation price reduction. The detailed objective function and constraints of this optimization is given in Appendix. Leveraging the Eastern Interconnection Planning Collaborative (EIPC) Phase I model that represents future uncertainties in EI generation, load, and transmission, PLEXOS modelling tool and commercial mix-integer programming (MIP) solver Xpress-XP were used to obtain the PV distribution at each PV penetration level. Additional inputs of this process included PV prices, land prices, as well as associated renewable portfolio standards, as shown in Table III. Based on the PV projection results, the PV penetration rate for each EI region in different scenarios is shown in Fig. 3. It is obvious that the regions in southern EI and those close to large load centers tend to have higher PV penetration rates. This is understandable because the southern EI regions have higher solar irradiance while the load center regions have higher local marginal prices and more economic surplus for more PV generation installation.

TABLE III
Data Resources for EI PV Projection

| PLEXOS model input | Data sources |
|---|---|
| Existing generation and transmission infrastructure, load forecast, solar radiation, fuel price forecast, carbon emission price forecast | The Eastern Interconnection Planning Collaborative (EIPC) database |
| PV price forecast | North American PV Outlook |
| PV sitting land price | Land Value 2015 Summary |

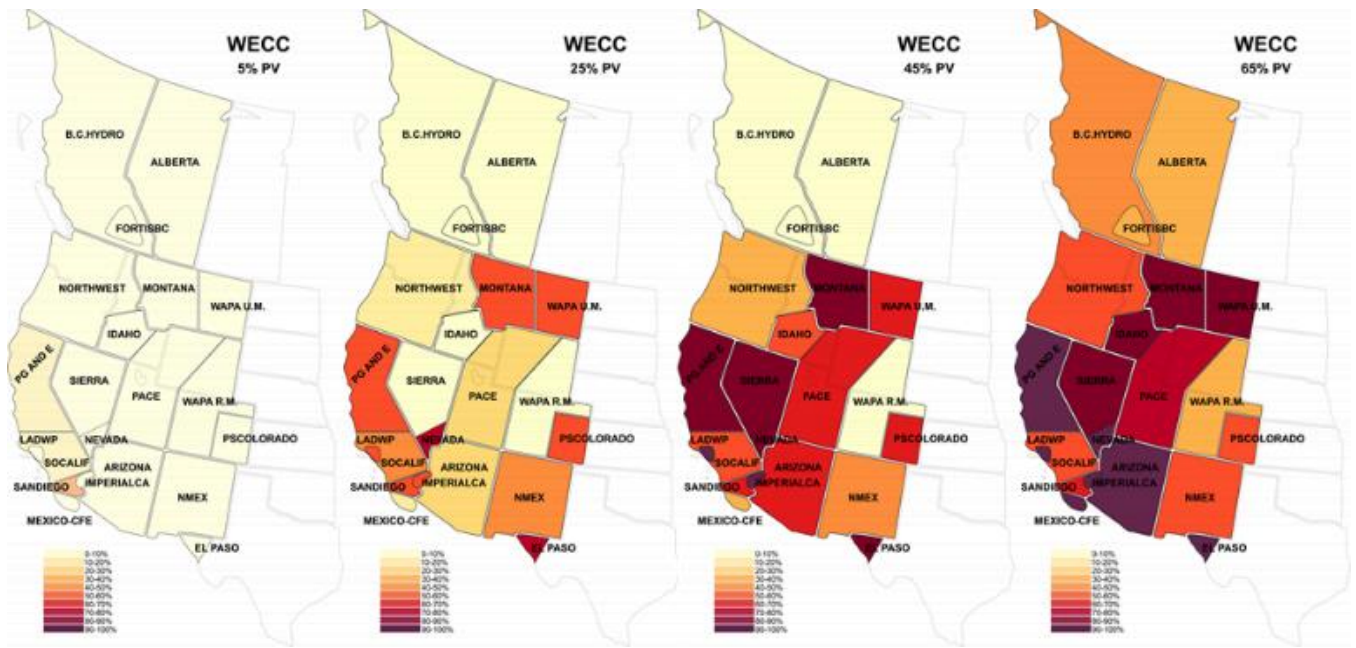

(a) 5% PV (b) 25% PV (c) 45% PV (d) 65% PV

Fig.4 PV penetration rate for each WECC region in different simulation scenarios

Regarding WECC, the data resources listed in Table IV were utilized to project the PV distribution for each WECC region. Specifically, the PV and wind distributions in the WECC 2022 light spring model and Renewable Portfolios Standard (RPS) 2022 data were used as the starting point. Then PV generation installation was proportionally distributed to each region according to the PV technical potentials estimated based on the National Solar Radiation Database (NSRDB). Some resource constraints and operational constraints were also considered, such as the maximum and minimum installed PV capacity in each region. The PV penetration rates for each WECC region is given in Fig. 4.

TABLE IV
Data Resources for WECC PV Projection

| Data Resource | Description |
|---|---|
| NSRDB data | Provide estimated PV power data based on average annual direct normal irradiance (DNI) |
| WECC 2022 LSP base case | Provide the load and generation capacity for different regions and the projected renewable location by 2022 |
| Renewable Portfolios Standard (RPS) 2022 data | Provide a RPS 2022 renewable portfolio by state, types, capacity, locations, and project process of renewable projects |

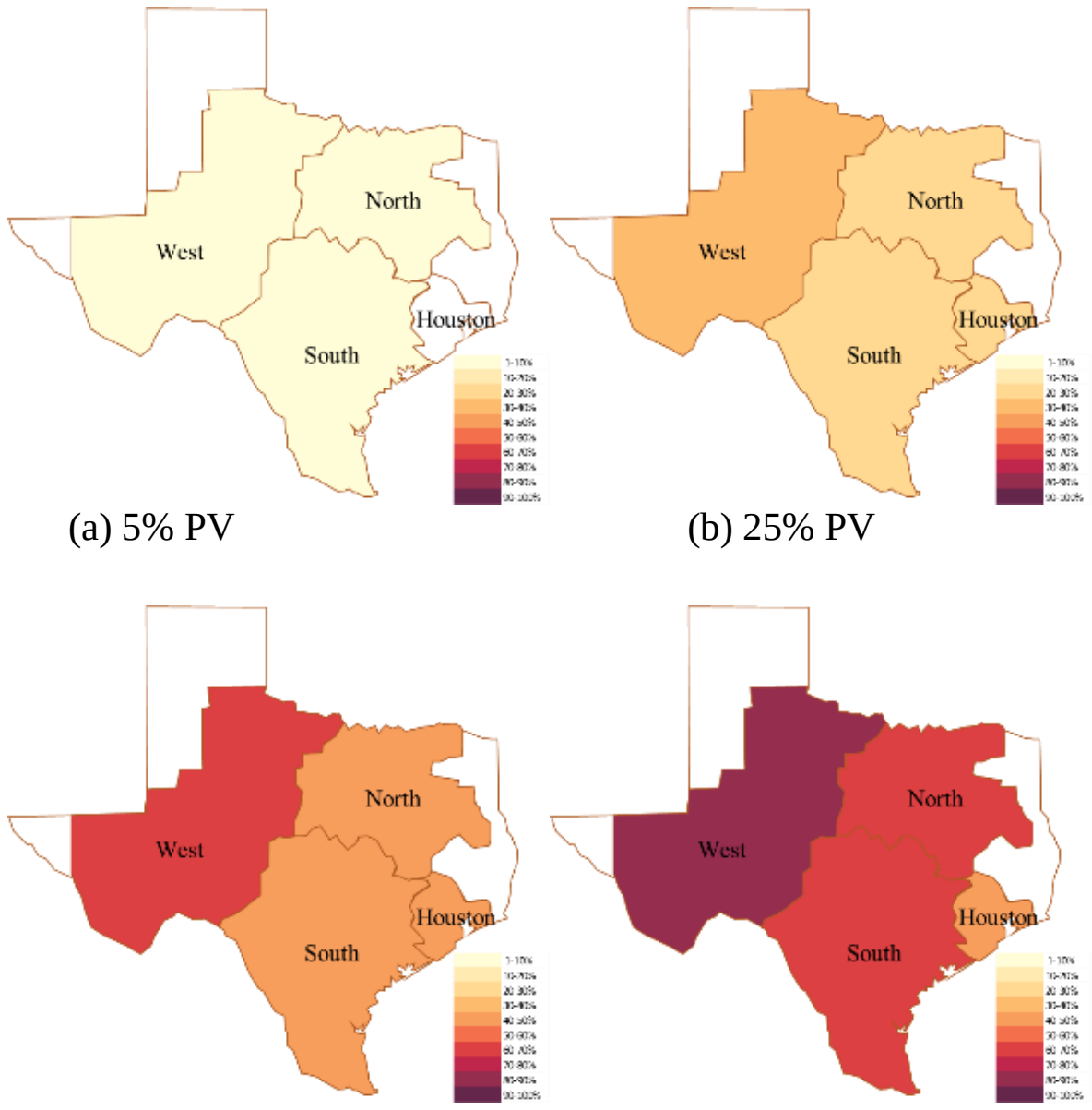

(a) 5% PV (b) 25% PV

(c) 45% PV (d) 65% PV

Fig.5 PV penetration rate for each ERCOT region in different simulation scenarios

ERCOT is a relatively small interconnection compared with EI and ERCOT and has only four regions. Firstly, the PV location information was used to obtain the PV distribution in the 5% PV penetration scenario, in which PV generation will be mostly located in the west region due to low land cost and rich solar irradiance. Then, PV generation in the west region was increased till this region's maximum generation capacity. Afterwards, PV generation was allocated to other three regions proportionally according to each region's average solar irradiance. The PV penetration rate for each ERCOT region is given in Fig. 5.

It is worth mentioning that the PV projection methodologies presented in this paper were merely the best practices that can be done based on available data sources, other than a perfectly accurate prediction. The TRC members agreed that the PV projection results should be accurate enough for interconnection level frequency response studies.

## *E. Dynamic Model Construction*

Leveraging the validated base models and projected PV geographic distribution, the dynamic models for each simulation scenario defined in Table II were constructed for each interconnection. This was a straightforward process but required large amounts of efforts to eliminate any numerical divergence issues.

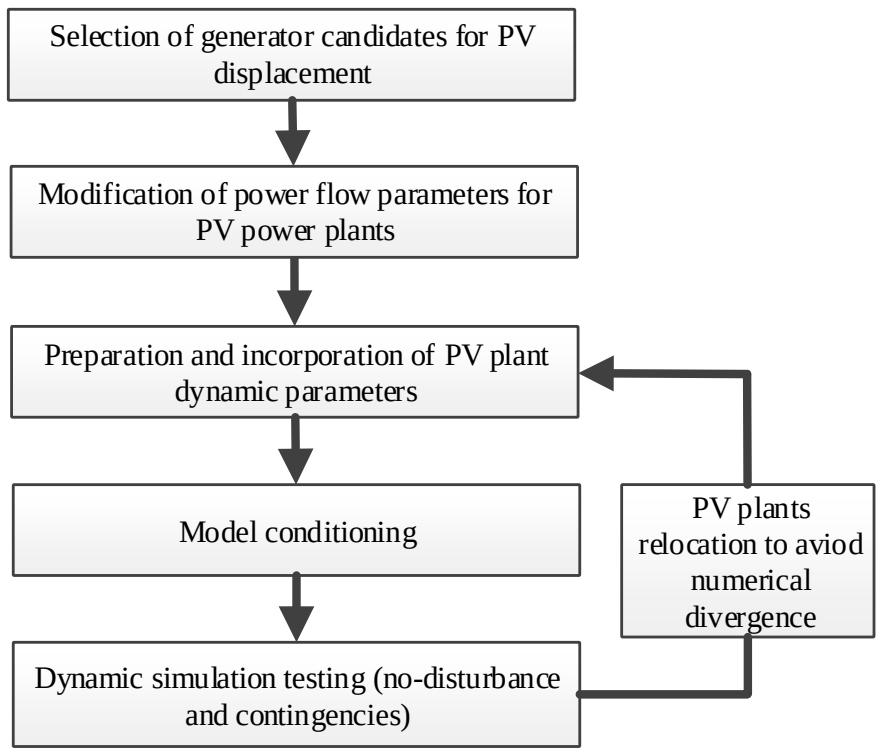

Fig.6 Flowchart of dynamic model construction for each simulation scenario

A flowchart of this process is given in Fig. 6. It is worth mentioning that all the models' frequency deviation is less than $10^{-5}$ pu during a 20-second no-contingency dynamic simulation run (also referred to as flat run). Contingency test runs were also carried out to examine the dynamic models' divergence conditions. Both no-contingency and contingency runs demonstrate that the constructed dynamic models are numerically stable and ready for frequency response studies.

## III. Impact on Interconnection Level Frequency Response

In order to evaluate the PV impact on three interconnections' frequency responses, dynamic simulations were conducted in this section using the dynamic models developed in previous sections. Both the largest contingency events recommended by NERC standard BAL-003-1 and typical events that occur on a daily basis were simulated.

### A. EI

To evaluate the EI frequency response deterioration due to increasing PV penetration, three generation loss contingencies with different magnitudes were simulated for EI. The first contingency was selected to be the largest contingency (4,500 GW generation loss) in the last 10 years while two additional contingency events included a typical N-2 contingency (2,250 MW generation loss) and a typical N-1 contingency (1,128 MW generation loss). These three contingencies are summarized in Table V and the frequency responses after these events in each simulation scenario are presented in Fig. 7. From Fig.7, it is easy to observe the EI frequency response deterioration with the increasing PV penetration.

TABLE V
Simulated Contingencies for EI

| Contingency | Description | Unit and Location | Capacity (MW) |
|---|---|---|---|
| 1 | The largest resource event in last 10 years | 5 units in south Indiana (August 4, 2007 Disturbance) | 4,500 |
| 2 | A typical N-2 contingency | 2 Braidwood Nuclear Units, Illinois | 2,250 |
| 3 | A typical N-1 contingency | 1 Browns Ferry Nuclear Unit, Alabama | 1,128 |

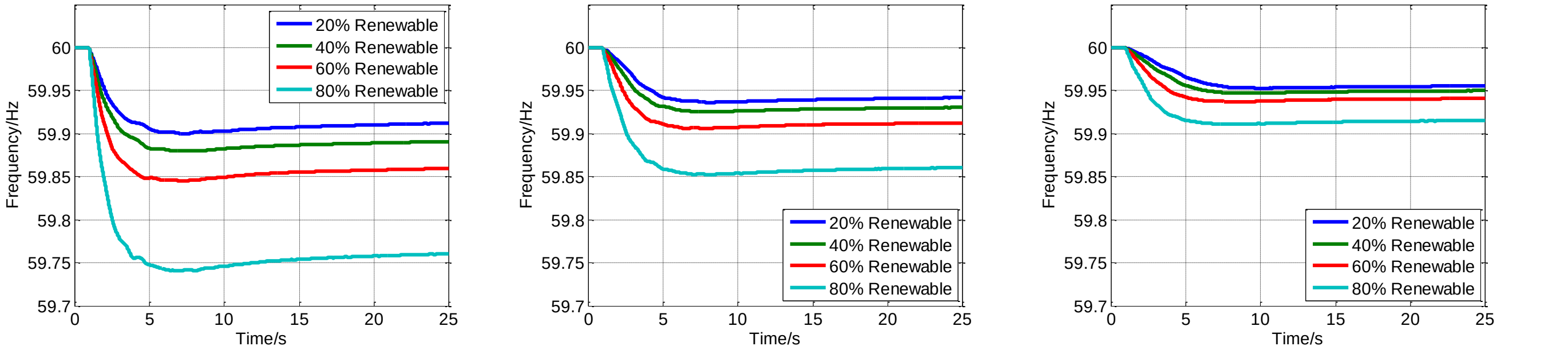

(a) 4,500 MW generation loss (b) 2,250 MW generation loss (c) 1,128 MW generation loss

Fig.7 EI frequency response simulation results in each simulation scenario

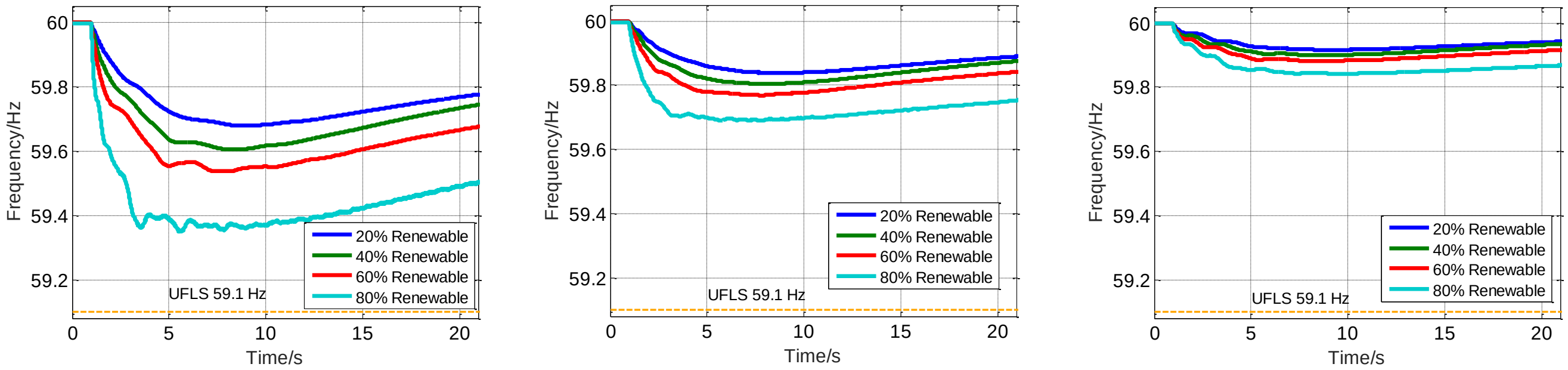

(a) 2,625 MW generation loss (b) 1,514 MW generation loss (c) 804 MW generation loss

Fig.8 WECC frequency response simulation results in each simulation scenario

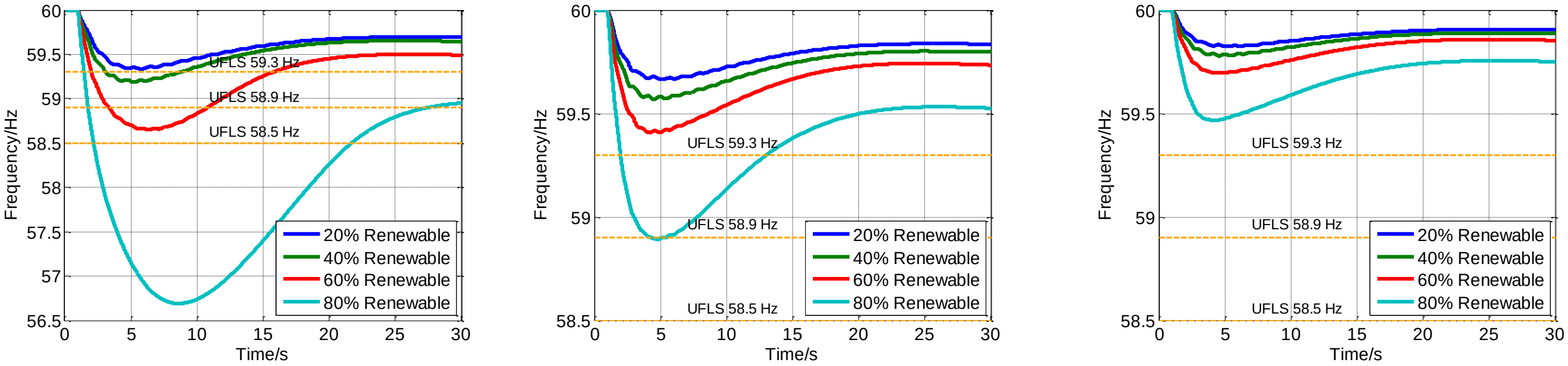

(a) 2,740 MW generation loss (b) 1,370 MW generation loss (c) 685 MW generation loss

Fig.9 ERCOT frequency response simulation results in each simulation scenario

## *B. WECC*

As shown in Table VI, the WECC contingency events included the largest N-2 contingency, a typical N-2 contingency, and a typical N-1 contingency. According to the WECC target resource contingency protection criteria, the loss of the two largest generating units in the Palo Verde nuclear power plant, totaling 2,625 MW, was chosen as the largest N-2 contingency. The loss of the two units in the Colstrip coal power plant, totaling 1,514 MW, and the loss of one unit in the Comanche power plant, totaling 804 MW, were chosen as the typical N-2 contingency and typical N-1 contingency, respectively.

The frequency responses of these three contingencies are shown in Fig. 8. The frequency responses of three contingencies have the same trend: the higher the PV penetration rate is, the worse frequency response becomes. This is obviously because both inertia and governor response decreases with the increasing renewable penetration rate.

TABLE VI
Simulated Contingencies for WECC

| Contingency | Description | Unit and Location | Capacity (MW) |
|---|---|---|---|
| 1 | The largest N-2 contingency | Two largest generating units in the Palo Verde nuclear plant | 2,625 |
| 2 | A typical N-2 contingency | Two units in the Colstrip coal power plant | 1,514 |
| 3 | A typical N-1 contingency | One unit in the Comanche plant | 804 |

## *C. ERCOT*

The contingency criteria recommended by NERC for ERCOT is the loss of two South Texas Project units carrying 2,750 MW load in total. In addition, a typical N-2 contingency (1,370 MW) and a typical N-1 contingency were also simulated (as shown in Table VII). ERCOT's frequency responses after these contingencies are shown in Fig. 9. As demonstrated by Fig. 9, despite the similar trend, ERCOT frequency response decreases much more dramatically than those of EI and WECC since ERCOT is a much smaller system.

TABLE VII
Simulated Contingencies for ERCOT

| Contingency | Description | Unit and Location | Capacity (MW) |
|---|---|---|---|
| 1 | The largest N-2 contingency | Two South Texas Nuclear Units | 2,740 |
| 2 | A typical N-2 contingency | Two Martin Lake Units | 1,370 |
| 3 | A typical N-1 contingency | One Martin Lake Unit | 685 |

## *D. NERC-defined Frequency Response Analysis*

In the NERC standard BAL-003-1, a metric was defined to evaluate each interconnection's frequency response and is calculated as the net change in generation divided by the change in frequency (in the unit of MW/0.1 Hz). This metric of each U.S. interconnection in each simulation scenario is presented in Fig. 10 and the NERC-recommended values for operation year 2017 are given in Table VIII.

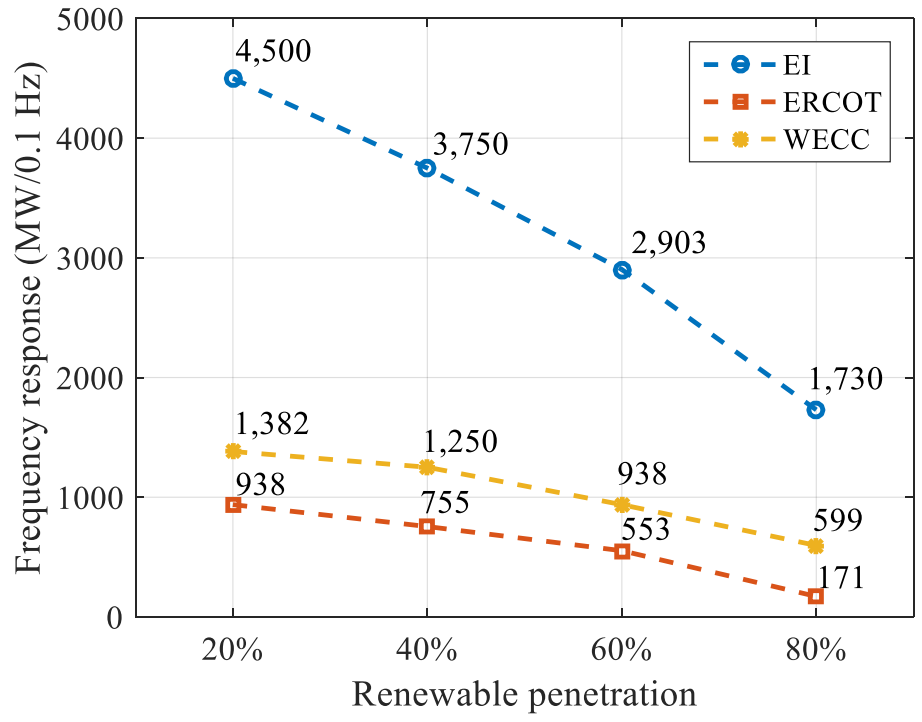

Fig.10 NERC-defined frequency response change for each interconnection

From Fig. 10, it is obvious that the NERC-defined frequency response metric declines with the increasing renewable penetration level for each interconnection, which is consist with the observations in the previous subsections. Furthermore, it demonstrates that despite the dramatic decline, the EI frequency response metric is still higher than the current NERC-recommended value even in the 80% renewable penetration scenario while WECC and ERCOT cannot comply with the current BAL-003-1 Standard in the same scenario without improving their frequency responses.

TABLE VIII
BAL-003-1 Recommended Frequency Response for Each Interconnection (2017)

| | EI | WECC | ERCOT |
|---|---|---|---|
| NERC-defined frequency response metric (MW/0.1Hz) | 1,015 | 906 | 471 |

## E. Under Frequency Load Shedding (UFLS) Analysis

The purpose of under frequency load shedding (UFLS) is to balance generation and load when a contingency causes a significant drop in frequency of an interconnection. It is a common technique to minimize the risk of system collapse and improve overall system reliability. Table IX presents a brief summary of the mainstream UFLS settings for each U.S. interconnection. Please note that the UFLS relay and under-frequency generator protection relay models were not included in the previous simulation runs in order to observe frequency nadirs clearly.

TABLE IX
Mainstream UFLS Settings for Each Interconnection

| | Load shedding block | Frequency set-point (Hz) | Percent of load dropped (%) | Cumulative percentage of load drooped (%) | Maximum trip Time (Cycles) |
|---|---|---|---|---|---|
| EI* | 1 | 59.5 | 6.5-7.5 | 6.5-7.5 | 18 |
| | 2 | 59.3 | 6.5-7.5 | 13.5-14.5 | 18 |
| | 3 | 59.1 | 6.5-7.5 | 20.5-21.5 | 18 |
| | 4 | 58.9 | 6.5-7.5 | 27.5-28.5 | 18 |
| ERCOT | 1 | 59.3 | 5 | 5 | 40 |
| | 2 | 58.9 | 10 | 15 | 40 |
| | 3 | 58.5 | 10 | 25 | 40 |
| WECC** | 1 | 59.1 | 5.3 | 5.3 | 14 |
| | 2 | 58.9 | 5.9 | 11.2 | 14 |
| | 3 | 58.7 | 6.5 | 17.7 | 14 |
| | 4 | 58.5 | 6.7 | 24.4 | 14 |
| | 5 | 58.3 | 6.7 | 31.1 | 14 |

*There are some minor regional variances. For example, Florida block 1 frequency set-point is 59.6 Hz.
**There are some slightly different sub-area plans in addition to the primary UFLS plan. For example, the block 1 frequency set-point of some sub-area plans is set to be 59.5 Hz.

As shown in Fig. 7, UFLS would not be triggered in EI even after the largest contingency (4,500 MW generation loss) in the 65% PV+15% WTG scenario. This means, despite the increasing PV penetration and decreasing frequency response, it is still unlikely for the EI system to resort to UFLS for frequency decline arresting. It is also worth mentioning that, it is common for system frequency to be lower than 60 Hz before a large contingency occurs in the real system. So NERC recommended the EI pre-contingency starting frequency to be 59.974 Hz for frequency response studies. Obviously, even taking into consideration this starting frequency bias, EI frequency nadir would be still above the UFLS frequency set-point. This mainly attributes to the EI system's large total inertia value and spinning reserve as an interconnected power grid. Regarding the WECC system, no primary UFLS would be triggered even in the 65% PV+15% WTG scenario after the largest N-2 contingency (2, 625 MW generation loss) neither. This observation is similar to that of the EI system. The difference is, since some sub-area UFLS set-point for WECC is 59.5 Hz, small amounts of load shedding would be possible in those areas, which could contribute to the WECC frequency response enhancement. Generally, the WECC system still has enough margin to stay away from large-scale UFLS at high PV penetration rates.

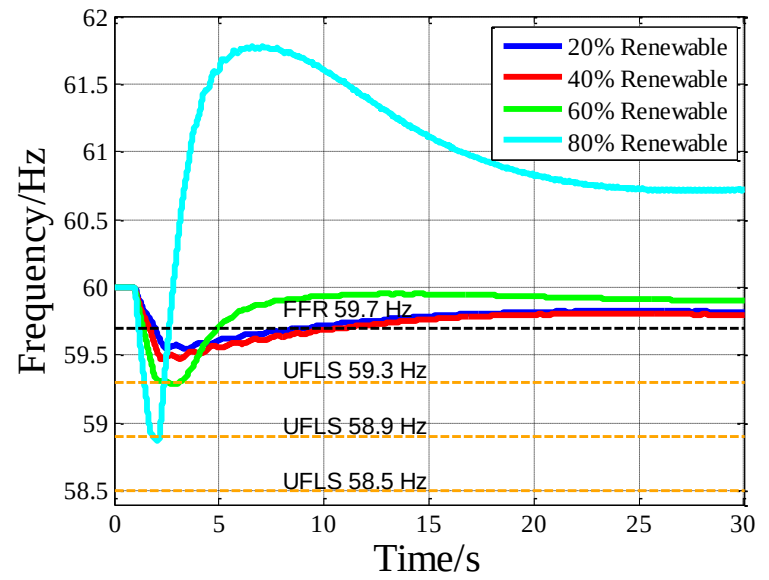

Fig.11 ERCOT frequency response with both FFR and UFLS relays after the largest N-2 contingency

Considering its much smaller size, it is not surprising for ERCOT to be much more vulnerable to UFLS especially compared with EI and WECC. This can also be easily observed from Fig. 9. That is why ERCOT starts to utilize a fast frequency response (FFR) scheme provided by fast responsive loads (e.g. large industrial loads, heat pumps, industrial refrigerator loads, and storage devices) as part of its frequency response enhancement efforts. This FFR scheme is actually a type of demand response. To provide FFR, a load resource will be equipped with an under-frequency relay (e.g. triggered if system frequency is dropped under 59.7 Hz). As required by ERCOT, the response time of FFR should be less than 30 cycles (including the frequency relay pickup delay and the breaker action time). Fig. 11 shows the simulation results when both FFR (1,400 MW load reduction at 59.7 Hz with a 30-cycle response time) and UFLS relays were simulated after the largest N-2 contingency (2,740 MW generation loss). It can be seen that FFR was triggered in all scenarios but first block UFLS (3,500 MW load reduction at 59.3 Hz with a 14-cycle tripping time) was triggered only in the 45% PV+15% WTG scenario and 65% PV+15% WTG scenarios. For the 45% PV+15% WTG scenario, since only some areas' frequency nadirs reached the first block UFLS set-point marginally, only 800 MW load was dropped, as shown in Table X. For the 65% PV+15% WTG scenario, dramatic frequency overshoot higher than 60 Hz were observed after first block UFLS action, which may trigger over-frequency generator protection relays and cause other stability issues. This overshoot is because, although only 1,300 MW (2,700 – 1,400) load shedding was needed after the 1,400 MW FFR to fully compensate the frequency decline, a total of 3,500 MW load was dropped by first block UFLS. These results indicate that ERCOT may need to tune its UFLS schemes in order to accommodate up to 80% renewables.

TABLE X
FFR and UFLS Amounts in Each ERCOT Scenario

| ERCOT Scenario | Generation loss Amount | FFR Amount | UFLS Amount/Percentage |
|---|---|---|---|
| 5% PV+15% WTG | 2.7 GW | 1.4 GW | 0 |
| 25% PV+15% WTG | | 1.4 GW | 0 |
| 45% PV+15% WTG | | 1.4 GW | 0.8 GW/1.14% |
| 65% PV+15% WTG | | 1.4 GW | 3.5 GW/5% |

## IV. Potential Mitigation Solutions

Countermeasures must be taken sooner or later in order to eliminate the negative effects of increasing PV penetration in the U.S. interconnections and prevent their frequency responses from rapid declining. In this section, the potential mitigation solutions are discussed.

### *A. Committing More Synchronous Generators*

To fight with the declining system inertia and frequency response, the most straightforward option for operators is to keep more synchronous generation units online through unit commitment. This will provide not only system inertia but also spinning reserve. But the drawbacks of doing this are obvious. Some of the synchronous generators will have to run at their minimum sustainable level, resulting in lower operation efficiency, more emission pollution, and higher energy price. PV and WTG generation production may also have to be curtailed in this situation. Furthermore, due to the minimum start-up time for synchronous generators, this option is not suitable for real time or urgent inertia compensation.

### *B. Adjusting Synchronous Generators' Governor Settings*

In order to take the most usage of the online synchronous generators available, their governor settings may be modified to help improve the system frequency response after generation loss contingencies. For example, as shown in Fig.12, employing a 3% governor droop instead of the original 5% droop will significantly improve the overall frequency response in the 5% PV +15% WTG scenario for all the interconnections.

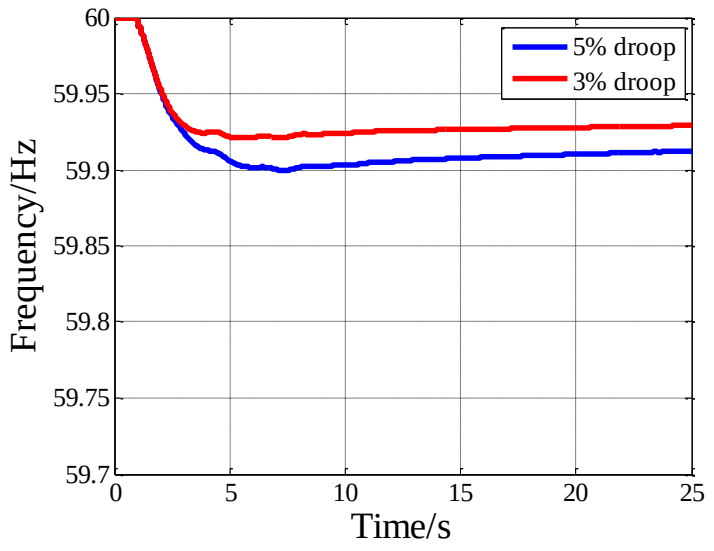

(a) EI

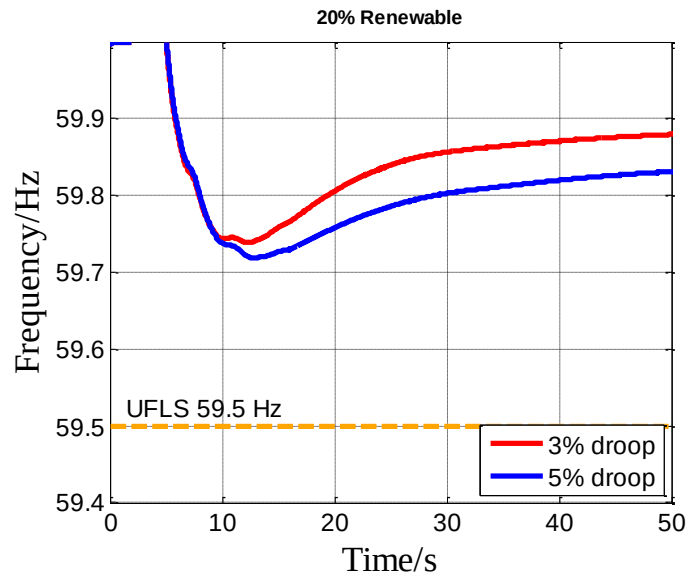

(b) WECC

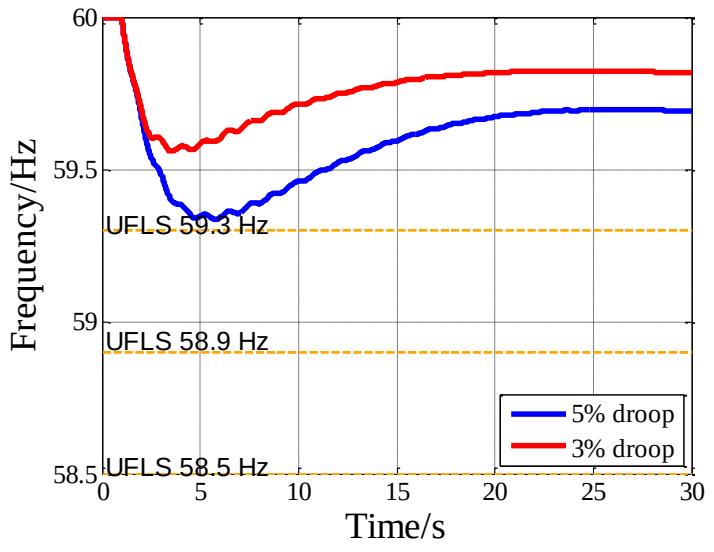

(c) ERCOT

Fig.12 Frequency responses with different governor droops in the 5% PV +15% WTG scenario

A 3% governor droop cannot provide more system inertia or spinning reserve but it leads to a faster governor response to severe under-frequency contingencies and thus a better primary frequency response. This allows the system to be operated under a lower inertia condition. Of course, the current governor droop is set to be 5% for good reasons and synchronous generators have their governor droop limits. Actually, if the droop is set to be lower than 3%, it may lead to low frequency oscillations and other system instability issues. So extensive contingency simulation studies have to be conducted before adjusting the synchronous generator governor droop universally in the real world.

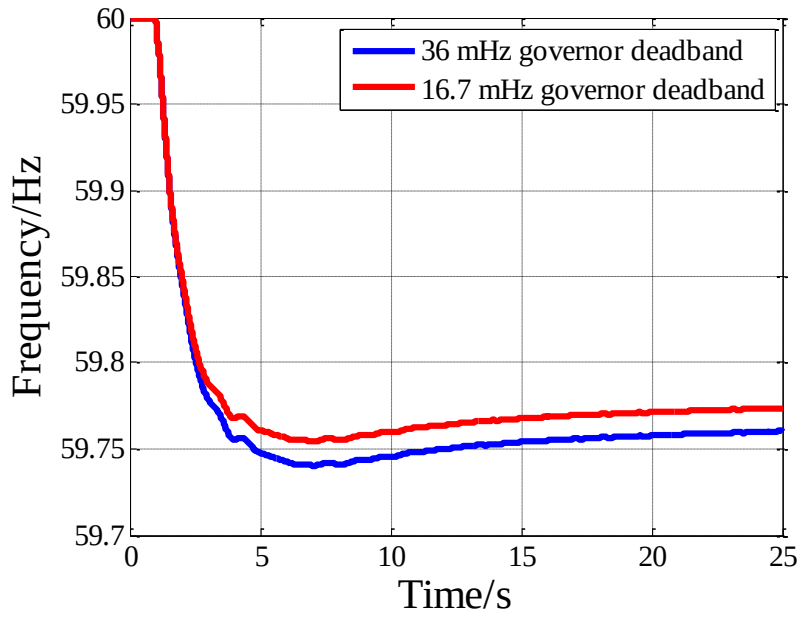

(a) EI

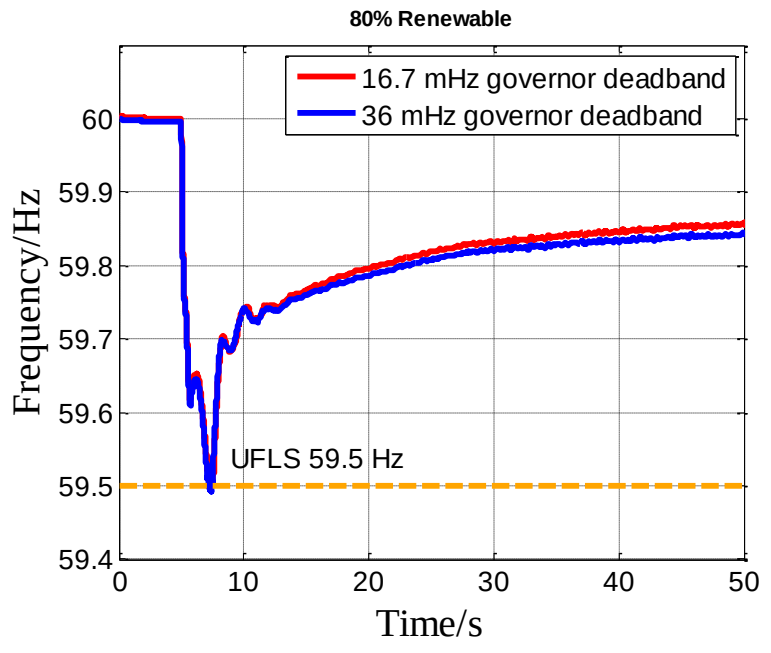

(b) WECC

Fig.13 EI and WECC frequency responses with different governor deadbands in the 65% PV +15% WTG scenario

Governor deadband is utilized to prevent excessive governor actions caused by small and frequent load variations but it also delay the governor responses to contingencies. So theoretically decreasing governor deadband should be able to let governors kick in earlier and help with overall frequency response. As shown in Fig. 13, decreasing the governor deadband from 36 mHz to the NERC-recommended 16.7 mHz, the EI and WECC frequency response was improved only slightly. This is because the overall frequency deviation caused by a large contingency is tens of times larger than governor deadband. So decreasing governor deadband does not have much potential to mitigate the frequency response decline.

### *C. Demand Response*

As mentioned earlier, ERCOT has been recruiting some load resources as a part of its frequency response reserve. If equipped with an under-frequency relay, load resources can provide full response in a few hundred milliseconds to under-frequency contingency events. Comparatively, a synchronous generator needs at least a few seconds to fully deliver its reserved capacity for frequency support due to the droop curve. This makes load response much more effective than synchronous generator governor response to improve frequency nadir when total system inertia is low and ROCOF is large. Fig. 14 shows the ERCOT frequency response improvement due to a 1.4 GW fast load response. Obviously, fast load response can be extremely valuable in preventing frequency from dropping below the involuntary UFLS threshold when the traditional spinning reserve is insufficient.

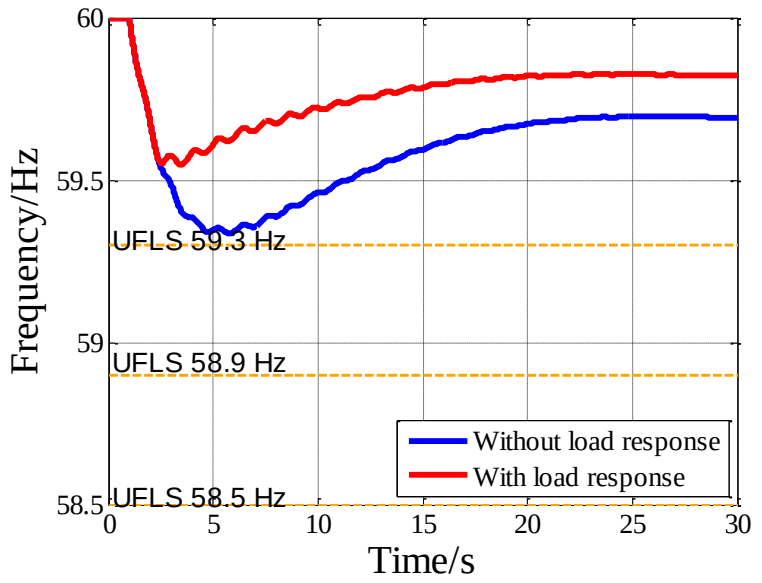

Fig.14 ERCOT frequency responses with fast load response in the 5% PV +15% WTG scenario

### D. *Synthetic inertia and governor response from converter-based resources*

Variable-speed WTGs, PV generations, and storage devices are interfaced with the synchronous grid via power electronic converters. Because of the fast response speed of power electronics devices, these converter-based resources are capable of regulating their power outputs in a fast manner. As shown by recent studies, these resources can be utilized to provide synthetic inertia and governor response for the purpose of fast frequency support. Particularly, required by the Standard Generation Interconnection Agreements (SGIA), the WTGs in the ERCOT shall provide primary frequency response similar to the droop characteristic of 5% used by conventional steam generators. Furthermore, the PV power plant's capability to provide similar response has been successfully demonstrated in the filed on a 300 MW PV power plant. Providing synthetic inertia is relatively complicated. The GE WindINERTIA technology enables a WTG to increase its power output by 5-10% of its rated turbine power to meet the inertial response grid requirement while no commercial PV inverter has been reported to have such functions. Generally, compared with synchronous generators, converter-based resources can be even more effective thanks to their fast power output regulation capabilities. Furthermore, it is understandable that a portion of their capacities have to be reserved for such services and some production curtailments are inevitable.

From the discussions above, almost all the frequency support options involve not only technical but also economic aspects. Besides verifying their effectiveness via dynamic simulations, a through cost-benefit analysis should be conducted to find the most cost-effective solution for different scenarios. Furthermore, ancillary service markets should be established to compensate those generation or load resources providing different fast frequency response services and the corresponding grid reliability requirements should also be updated.

## V. Conclusions

Extra-high PV penetration rates are expected to occur in the U.S. interconnections in the next several decades. The impact of extra-high PV penetrations on three different U.S. interconnections were evaluated in this paper using measurement-validated models and realistically-projected PV geographic distribution information. Potential solutions were also implemented and discussed to evaluate their effectiveness. The simulation results revealed that the U.S. interconnections are able to accommodate the high PV penetrations predicted by the DoE Sunshot vision study in terms of frequency response, although some additional frequency support regulations need to be engaged. This paper provides a guidance for future PV penetration in the large interconnections.

## VI. Appendix

### A. *Objective function*

The objective is to minimize the sum of all costs during the planning horizon. The objective function consists of three cost items: expansion cost, system operation cost, and emission cost. The expansion cost includes the PV panel price and the land cost. The operation cost is the sum of the fixed operation cost, the varying operation cost, the maintenance cost, the penalty of lost load, and the wheeling cost of the transmission network between balance authorities. The objective function is described in (1).

$$f = \sum_{y=1}^{N_Y} D_y \cdot \left\{ \begin{array}{l} C_{\text{PV expansion cost}} + C_{\text{fixed O\&M cost}} + C_{\text{varying O\&M cost}} \\ + C_{\text{fuel cost}} + C_{\text{wheeling cost}} + C_{\text{cost of lost load}} \\ + C_{\text{emission cost}} \end{array} \right\} \quad (1)$$

where $D_y$ is the coefficient of cash flow versus present value considering "the end year effect".

In this objective function, the PV expansion cost includes the PV panel installation costs and land costs. The PV panel prices is from predicted values and the land prices are differentiated for each region. For each generation technology type, the fixed O&M cost is determined by the rated MW capacity of a unit. The cost of fuel depends on the unit heat rate, fuel price and unit output. The varying O&M cost represents the incremental cost of machine wear and replacement associated with each MWh energy. The penalty of unserved load depends on the unserved energy and the penalty price of lost load in each region. The wheeling cost of interfaces between balancing authorities is calculated by the wheeling prices and energy flows on interfaces. The emission cost is obtained by the emission price and the generation emission coefficient of each unit of fossil power plants.

### B. *Chronology Modeling and Constraints*

To reduce the number of modeling intervals, the hourly variation of load and renewable resources in each year is converted into multiple time blocks, each of which represents a scenario of these stochastic variables. The length of each time block represents the duration of this operation scenario over each year. The PV allocation is obtained by minimizing the objective function (1) subject to following constraints.

1) Regional power balance constraint. In each time block, the generation, load, and the interface exchange in each region should be balanced.

2) PV installation speed constraint. The annual expansion of PV in each region is constrained by available resources and the grid integration approval processing.

3) Unit capacity constraint considering outages and scheduled maintenance discount. Due to the forced outages and scheduled maintenance, the available capacity will be smaller than the installed capacity for reliability and adequacy consideration.

4) Capacity adequacy constraint. The available generation capacity in each region should be able to provide load and reserve. The amount of reserve should be able to meet the requirement of interface and unit contingencies.

5) Interface transmission capacity constraint. Interface power flow should be within its limit.

6) Regional renewable portfolio constraint. The renewable generation in some regions are required to be higher than a percentage of the total regional generation.

7) PV plants output constraint. The maximum output of PV plants is constrained by the availability of the solar radiation, represented by a fix value for each time block for each region. This fix radiation value is converted from time-series solar radiation data during scenario creation.

The optimization problem to minimize (1) subject to constraint 1)-7) forms a mixed integer programming problem and can be solved by commercial MIP solvers. The PLEXOS tool is applied to formulate and solve this problem.